\documentclass[twocolumn,twocolappendix]{aastex701}

\usepackage{amsmath}
\usepackage{graphicx}
\usepackage{soul}
\usepackage{hyperref}

\definecolor{red}{RGB}{190,0,0}

\definecolor{pink}{RGB}{255,51,153}

\newcommand{\msun}{\mbox{$M_{\odot}$}}
\newcommand{\msunyr}{\mbox{$M_{\odot}\,\mathrm{yr^{-1}}$}}
\newcommand{\kms}{\mbox{$\mathrm{km\,s^{-1}}$}}

\newcommand{\rvir}{\mbox{$R_\mathrm{vir}$}}
\newcommand{\mstar}{\mbox{$M_{\star}$}}

\newcommand{\nH}{\mbox{$n_\mathrm{H}$}}
\newcommand{\HI}{\mbox{\ion{H}{1}}}
\newcommand{\Lya}{Ly{$\alpha$}}
\newcommand{\fLyC}{\mbox{$f_{900}$}}
\newcommand{\vsep}{\mbox{$v_\mathrm{sep}$}}

\newcommand{\fLya}{\mbox{$f_{\mathrm{Ly}\alpha}$}}
\newcommand{\CRMHD}{\texttt{CRMHD}}
\newcommand{\MHD}{\texttt{MHD}}
\newcommand{\BR}{\mbox{$B/R$}}
\newcommand{\vred}{\mbox{$v_\mathrm{red}$}}
\newcommand{\fvalley}{\mbox{$f_\mathrm{valley}$}}

\received{\today}
\revised{\today}

\shorttitle{Lyman-$\alpha$ with CR-driven outflows}
\shortauthors{Taysun Kimm et al.}

\begin{document}

\title{Impact of Cosmic Ray-driven Outflows on Lyman-$\alpha$ Emission in Cosmological Simulations}

\author[0000-0002-3950-3997]{Taysun Kimm}
\affiliation{Department of Astronomy, Yonsei University, 50 Yonsei-ro, Seodaemun-gu, Seoul 03722, Republic of Korea}
\email[show]{tkimm@yonsei.ac.kr}

\author[0000-0002-8140-0422]{Julien Devriendt}
\affiliation{Astrophysics, University of Oxford, Denys Wilkinson Building, Keble Road, Oxford OX1 3RH, UK}
\email{julien.devriendt@physics.ox.ac.uk}

\author{Francesco Rodr\'{\i}guez Montero }
\affiliation{Department of Astronomy \& Astrophysics, University of Chicago, 5640 S Ellis Avenue, Chicago, IL 60637, USA}
\affiliation{Astrophysics, University of Oxford, Denys Wilkinson Building, Keble Road, Oxford OX1 3RH, UK}
\email{currodri@uchicago.edu}

\author{Adrianne Slyz}
\affiliation{Astrophysics, University of Oxford, Denys Wilkinson Building, Keble Road, Oxford OX1 3RH, UK}
\email{adrianne.slyz@physics.ox.ac.uk}

\author[0000-0002-7534-8314]{J\'er\'emy Blaizot}
\affiliation{Univ Lyon, Univ Lyon1, Ens de Lyon, CNRS, Centre de Recherche Astrophysique de Lyon UMR5574, F-69230 Saint-Genis-Laval, France}
\email{jeremy.blaizot@univ-lyon1.fr}

\author[0000-0003-1561-3814]{Harley Katz}
\affiliation{Department of Astronomy \& Astrophysics, University of Chicago, 5640 S Ellis Avenue, Chicago, IL 60637, USA}
\affiliation{Kavli Institute for Cosmological Physics, University of Chicago, Chicago IL 60637, USA }
\email{harleykatz@uchicago.edu}

\author[0009-0002-3737-8384]{Beomchan Koh}
\affiliation{Department of Astronomy, Yonsei University, 50 Yonsei-ro, Seodaemun-gu, Seoul 03722, Republic of Korea}
\email{bckoh@yonsei.ac.kr}

\author[0000-0002-4362-4070]{Hyunmi Song}
\affiliation{Department of Astronomy and Space Science, Chungnam National University, Daejeon 34134, Republic of Korea }
\email{hmsong@cnu.ac.kr}

% Abstract of the paper
\begin{abstract}
Cosmic ray (CR) feedback has been proposed as a powerful mechanism for driving warm gas outflows in galaxies. We use cosmological magnetohydrodynamic simulations to investigate the impact of CR feedback on neutral hydrogen (\HI) in a $10^{11}\,\msun$ dark matter halo at $2<z<4$. To this end, we post-process the simulations with ionizing radiative transfer and perform Monte Carlo Lyman-$\alpha$ (\Lya) transfer calculations. CR feedback reduces \HI\ column densities around young stars, thereby allowing more \Lya\ photons to escape and consequently offering a better match to the \Lya\ luminosities of observed \Lya\ emitters. Although galaxies with CR-driven outflows have more extended \HI\ in the circumgalactic medium (CGM), two \Lya\ line properties sensitive to optical depth and gas kinematics -- the location of the red peak in velocity space ($\vred$) and relative strength of the blue-to-red peaks ($\BR$) -- cannot distinguish between the CR-driven and non-CR simulations. This is because \Lya\ photons propagate preferentially along low \HI\ density channels created by the ionizing radiation, thereby limiting the scattering with volume-filling \HI. In contrast, the observed low flux ratios between the valley and peak and the surface brightness profiles are better reproduced in the model with CR-driven outflows because the \Lya\ photons interact more before escaping, rather than being destroyed by dust as is the case in the non-CR simulation. We discuss the potential cause of the paucity of sightlines in simulations that exhibit prominent red peaks and large \vred, which may require the presence of more volume-filling \HI. 
\end{abstract}


\section{Introduction}

\setcounter{footnote}{0}

Galaxy formation models require strong gaseous outflows to explain inefficient star formation observed in the local Universe \citep[e.g.,][]{naab17}. However, the driving mechanisms of these galactic outflows remain unclear. While the explosion of massive stars is considered a powerful regulator of star formation, numerical simulations have often struggled to reproduce observed galaxy properties, such as the stellar mass-to-halo mass ratio, without the explosion energy being artificially boosted \citep[][c.f., \citet{hopkins18}]{li18,rosdahl18}. Although the inclusion of UV radiation feedback can potentially lower the density into which supernova (SN) remnants propagate \citep{geen15}, radiation alone is insufficient to drive outflows exceeding several hundreds of \kms, as the neutral medium is collisionally ionized rapidly before it is accelerated. Non-thermal pressure from multi-scattered infrared \citep[e.g.,][]{murray05,hopkins11} or Lyman-$\alpha$ (\Lya) photons \citep[e.g.,][]{smith17,kimm18,nebrin25} is also suggested as an effective mechanism to launch winds in dusty or dust-free regions, respectively. However, the number of scatterings is highly sensitive to the medium's geometry and velocity structures \citep{skinner15,gronke16a}, thereby rendering the non-thermal radiation pressure a less robust solution to the overcooling problem, which occurs in diverse environments. It is also possible that multiple processes operate simultaneously, but the effects of stellar feedback generally become weakened with increasing density \citep[e.g.,][]{kimjg23}, thus making it susceptible to radiative losses. 

To prevent runaway cooling and regulate star formation, an energy source  effective at high densities ($\nH\gtrsim100\,\mathrm{cm^{-3}}$) is required. One promising candidate is the non-thermal pressure support from cosmic rays (CRs), which can be generated across shocks driven by SNe or active galactic nuclei via diffusive shock acceleration. Owing to its relativistic nature, CR pressure decays more  slowly than does the thermal pressure of an ideal gas. The additional support from CRs in the galactic midplane is known to drive gentle, warm outflows with temperatures on the order of $10^{4-5}\,\mathrm{K}$ \citep{girichidis16,ruszkowski17}, which are absent in the canonical interstellar (ISM) model, wherein thermal instability rapidly removes such gas. Early studies assuming a constant CR diffusion coefficient show that stronger outflows reduce the star formation rates by a small factor in isolated disk galaxies \citep{salem14,chan19,dashyan20,semenov21,farcy22}. \citet{farcy25} further show that gas displaced by CR feedback fills the low-density channels, thereby increasing the optical depth to Lyman continuum (LyC) photons and thus delaying reionization of the universe. A natural question arises as to whether CR pressure leaves observable traces in galaxies.

Historically, the stellar mass-to-halo mass relation, shape of galactic rotation curves, and chemical information have been useful probes of galaxy formation \citep[to name a few,][]{hopkins18,dave19,nelson2019_tng50,agertz21,dubois21}. Confronting the mass loading of galactic outflows is another useful approach \citep{veilleux05,pandya21}; however, detailed modeling of spectroscopic line formation is required for a direct comparison with observations \citep{hummels17,mauerhofer21}. Alternatively, spectroscopic signatures in quasar absorption lines can be used to infer hydrogen or metal column densities in the circumgalactic medium (CGM) \citep{werk13,prochaska17}, constraining stellar feedback models. \citet{DeFelippis24} show that although still insufficient, a CR model with a high diffusion coefficient $\kappa=3\times10^{29}\,\mathrm{cm^2\,s^{-1}}$ can account better for the large \ion{C}{4} column densities observed.

Alongside the absorption features, \Lya\ emission provides a unique window through which the physical properties of the neutral medium can be studied. Owing to its resonant nature, \Lya\ emission forms a double-peaked line profile, with peak velocities mainly determined by the optical depth \citep{dijkstra14,lao20}, and the separation of the double peak $v_\mathrm{sep}$ given by 
\begin{equation}
v_\mathrm{sep}=336\,\kms \, \sqrt{1+\mathcal{M}^2} \, N_{20}^{1/3} \, T_4^{1/6},
\label{eq:vred}
\end{equation}
where $\mathcal{M}$ is the Mach number, $N_{20}\equiv N_\mathrm{HI}/10^{20}\,\mathrm{cm^{-2}}$, $N_\mathrm{HI}$ is the \HI\ column density, and $T_4\equiv T/10^4\,\mathrm{K}$, in the case of a dust-free uniform sphere.
Moreover, the relative flux ratio between the blue and red sides of the line profile (\BR) can provide information about the bulk motion of the scattering medium, as backscattering photons enhance the flux redward (or blueward) of the line centre depending on whether the gas is outflowing (or inflowing) \citep{dijkstra06,verhamme06}. Analyzing the \Lya\ spectrum of star-forming galaxies, \citet{trainor15}, for example, found that, at $z\sim2$, the majority of actively star-forming galaxies exhibit $\BR<1$, suggesting that the scattering medium around the galaxies is outflowing; this is consistent with the presence of blueshifted low-ionization metal-absorption lines \citep{steidel10}. The ubiquitous detection of large \Lya\ halos around high-$z$ star-forming galaxies is also noteworthy \citep{rauch08,steidel11,hayes13,wisotzki16,leclercq17,wisotzki18}. Using a simple 1D halo model with varying densities and velocities, \citet{song20} demonstrate that both \Lya\ lines and surface brightness (SB) profiles of some of the \citet{leclercq17} samples can be recovered if the halo has an extended neutral outflow. \citet{Erb2023} further show that not only the \Lya\ properties but also the low-ionization metal absorption lines of twelve $10^9\,\msun$ galaxies at $z\sim2$ are well fitted by multiphase clumpy outflow models \citep[see also][]{Garel24}. 

Because an extended CGM constitutes one of the most distinctive features of  simulations with CR-driven outflows, it is interesting to investigate whether the predicted \Lya\ lines are significantly affected by CRs. Using stratified ISM simulations \citep{girichidis18}, \citet{gronke18} argue that the two characteristic features of Lyman alpha emitters (LAEs) -- strongly red-peaked \Lya\ lines with minimal flux at the line centre, are better reproduced when CR pressure is included. However, the large value of the location of the red peak in velocity space (\vred) is not reproduced even when CR-driven outflows are present. Given that the setup of these simulations is idealized, revisiting this issue by analyzing galaxies simulated in a cosmological environment, where discontinuous external accretion events can drive stronger outflows, is necessary. Recently, post-processing a cosmological simulation of a halo of mass $5\times10^9\,\msun$ at $z=3.5$, \citet{yuan24}  demonstrated that a dwarf galaxy simulated with CRs exhibits a more extended \Lya\ halo than the one without, thereby hinting at the importance of CR driven outflows in \Lya\ prediction. We build upon these studies and use recent cosmological magnetohydrodynamic simulations of a more massive halo ($10^{11}\,\msun$ at $z\sim2$), run with and without CRs \citep{rodriguez-montero24}, to investigate the differences in \Lya\ features. We note that the simulated galaxies differ significantly in stellar mass and CGM structures, depending on the presence of CRs; this provides an interesting testbed to investigate if \Lya\ can be used to constrain galaxy formation models. 

The remainder of this paper is structured as follows: Section~\ref{sec:methods} describes the simulations and post-processing methods used to generate the mock \Lya\ spectra and images. In Section~\ref{sec:results}, we present our analysis of the \Lya\ properties of a simulated galaxy at different redshifts and their comparison with observed LAEs. In Section~\ref{sec:discussion}, we compare the simulation results with and without the ionizing radiative transfer and explore the implications and potential caveats of our interpretations. Finally, we conclude this paper with a summary and future prospects in Section~\ref{sec:conclusions}.

\section{Methods}
\label{sec:methods}

\subsection{Magnetohydrodynamic simulations}
To study \Lya\ line properties of high-$z$ galaxies, we use  cosmological magnetohydrodynamic simulations run with the adaptive mesh refinement (AMR) code {\sc ramses} \citep{teyssier02,fromang06}. The simulations use the {\sc Nut} galaxy zoom-in initial condition \citep{powell11}, which contains a  $\sim 10^{11}\,\msun$ dark matter halo at $z\sim2$. Gas cools via atomic and metallic species, depending on the gas-phase metallicity, with a uniform UV background turned on at $z=9$ \citep{haardt96}. Star particles form in gravitationally unstable regions with an efficiency set by the local magneto-thermo-turbulent conditions \citep{kimm17,martin-alvarez20}. Type II SNe of an explosion energy of $10^{51}\,\mathrm{erg}$ are modeled using the mechanical feedback scheme \citep{kimm15}, with a rate of 0.02 $\msun^{-1}$. We assume that 10\% of the SN energy is transferred to the accelerated CRs. Our fiducial run includes not only anisotropic diffusion of CRs with a coefficient of $\kappa=3\times10^{28}\,\mathrm{cm^{2}\,s^{-1}}$, but also streaming at the local Alfv\'en wave velocity \citep{dubois19}. The simulations employ an initial magnetic field of $3\times10^{-12}\,\mathrm{G}$, which amplifies to a few $\mu \mathrm{G}$ within a Gyr timescale. The dark matter and stellar mass resolution are $5.5\times10^4\,\msun$ and $4.5\times10^3\,\msun$, respectively, and the maximum AMR resolution is $\approx$23 pc (physical). A more detailed explanation of the simulations can be obtained from \citet{rodriguez-montero24}.

\subsection{Lyman-$\alpha$ radiative transfer}

To compute the \Lya\ emission, we use the Monte Carlo radiative transfer (MCRT) code {\sc rascas} \citep{michel-dansac20}. A recombinative source term is calculated from the spectral energy distribution (SED) of each star particle\footnote{In principle, the recombinative term can also be obtained from the ion species, but we adopt this approximation to facilitate the direct comparison between simulations with and without post-processing of ionizing radiation transport.}, while the abundance of hydrogen ion species from the simulations is used to calculate a collisional term (if necessary). We use the Binary Population and Spectral Synthesis code \citep[v2,][]{stanway16} to obtain the SED of each stellar particle, which is characterized by its stellar mass, metallicity, and age. Because only a small fraction of ionizing photons tends to escape from star-forming galaxies at high redshift \citep[e.g.,][]{Saxena22}, we assume that all the LyC photons are absorbed by nearby neutral hydrogen, and 68\% are reprocessed into \Lya.

Because the cosmological simulations are performed without fully-coupled ionizing radiation transport, we post-process the simulation outputs using {\sc ramses-rt} \citep{rosdahl13}, and update the ionization and temperature structures (see Appendix~\ref{sec:appendix}). This significantly alters our interpretation of the results, as discussed below.  For snapshots processed with {\sc ramses-rt}, we additionally account for \Lya\ produced via collisional excitation, as in \citet{kimm22}. Because the resolution of the simulations is finite, we ignore collisional excitation from cells, wherein the local cooling time ($t_\mathrm{cool}$) is underesolved, i.e. $\Delta t > t_\mathrm{cool}/10$ \citep{mitchell21}; here, $\Delta t$ is the local simulation (fine) timestep. \Lya\ photons are then propagated, interacting with neutral hydrogen and deutrium with a ratio of $D/H=3\times10^{-5}$ until they escape from the virial sphere or are destroyed by dust. We adopt a simple dust prescription, in which dust is assumed to be present in the neutral gas, with only 1\% surviving in the fully ionized media \citep{laursen09}. The amount of dust is scaled to the metal mass using the dust-to-metal ratio for the Small Magellanic Cloud, inferred from the METAL program \citep{roman-duval22}. A uniform turbulence of 20\,\kms\ is assumed, when calculating the \Lya\ interaction probabilities. We use $10^5$ photon packets by default and re-run the MCRT until at least 2000 photon packets contribute to the angle-averaged emission line profile. We also generate 10 mock spectra along randomly chosen sightlines for a more direct comparison with the observations. To define the flux redward ($R$) or blueward ($B$) of the \Lya\ spectrum, we compute the attenuated $\mathrm{H}\alpha$ emission based on stellar particles for each angle-averaged or mock spectrum and take its flux average as the line center.

\begin{figure}
	\includegraphics[width=\columnwidth]{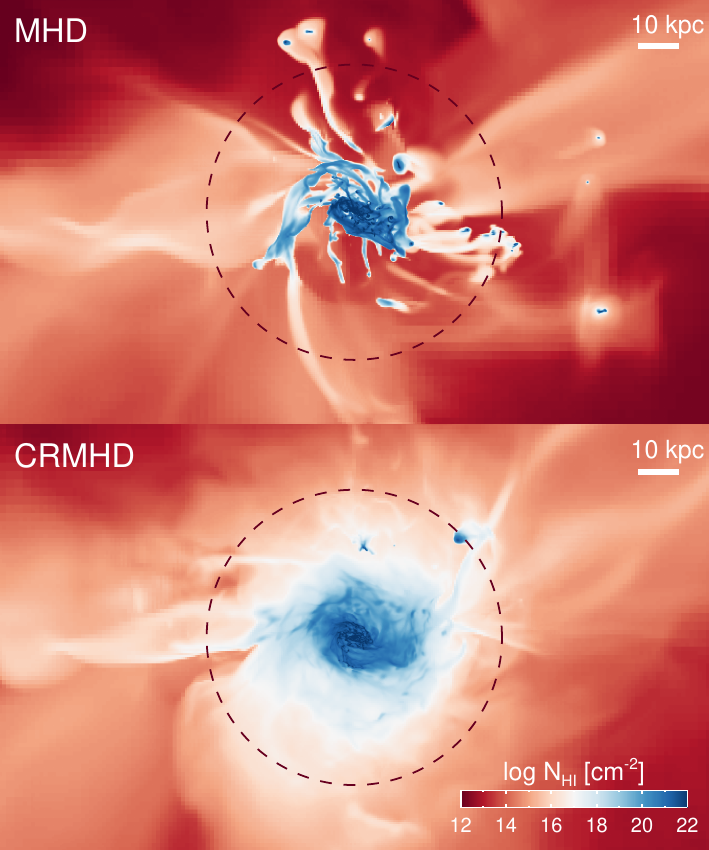}
    \caption{Distribution of neutral hydrogen column density ($N_\mathrm{HI}$) for the simulated halo with $3\times10^{11}\msun$ at $z\approx 2$ from the runs without and with CRs (top and bottom, respectively). The dashed circles denote $0.5\, R_\mathrm{vir}$, where $R_\mathrm{vir}$ is the virial radius of the dark matter halo ($\approx 72\,\mathrm{kpc}$).  The neutral hydrogen fraction is re-computed by post-processing each snapshot with {\sc ramses-rt}. The galaxy simulated with CR feedback clearly has a more extended HI distribution than the one without.}
    \label{fig:img}
\end{figure}

\section{Results}
\label{sec:results}

\subsection{Global galactic properties}
The average stellar mass of the simulated galaxy is $\mstar\approx 3\times10^{9}\,\msun$ in the fiducial run with CR feedback at $2\lesssim z \lesssim 4$. As discussed already in \citet{rodriguez-montero24}, this is in reasonable agreement with that predicted from the stellar-to-halo mass relation derived from the halo abundance matching technique \citep[e.g.,][]{behroozi13}.  The simulated galaxy is rotationally supported, with an average star formation rate of $\sim4\,\msunyr$ and a stellar metallicity of $\approx 0.25\, Z_{\odot}$. The dust-attenuated UV magnitude at $1500\,\AA$ is  $M_{1500}\sim -18.5$, and the UV slope measured at $1268 \le \lambda/\,\AA \le 2580$ \citep[e.g.,][]{calzetti94} is $\beta_{UV}\approx-1.5$. Note that these are similar to the typical properties of the LAEs at $z\sim3$, although the \CRMHD\ galaxy may be at the slightly more massive end of these LAEs \citep[e.g.,][]{gawiser07}. In contrast, the \texttt{MHD} run (i.e. galaxy simulated without CRs) is four times more massive ($\mstar\approx 1.4\times10^{10}\,\msun$), metal-rich ($\approx 0.63\, Z_{\odot}$), and more actively star-forming ($\sim8\,\msunyr$), but redder with $\beta_{UV}\approx-1.2$. At $z=2$,  the simulated galaxy in the MHD run has a massive stellar core at the center with the maximum circular velocity reaching 370\, \kms. All of these features suggest that the \texttt{MHD} run suffers from overcooling \citep[e.g.,][]{kimm15}, whereas the star formation is more controlled in the \CRMHD\ run. These two simulations, therefore, constitute a useful set to investigate whether different physical processes leave any distinct imprints on the \Lya\ observables. We reemphasize that this comparison is particularly intriguing because CR feedback is known to drive warm outflows, which can give rise to more extended neutral hydrogen  \citep[e.g.,][]{girichidis16,farcy22}. Figure~\ref{fig:img} indeed shows that the galaxy simulated with CRs exhibits an extended CGM with $N_\mathrm{HI}\sim10^{17-18}\,\mathrm{cm^{-2}}$, whereas, in the \MHD\ run, most of this gas tends to collapse directly onto the central galaxy.

% Example figure
\begin{figure}
	\includegraphics[width=\columnwidth]{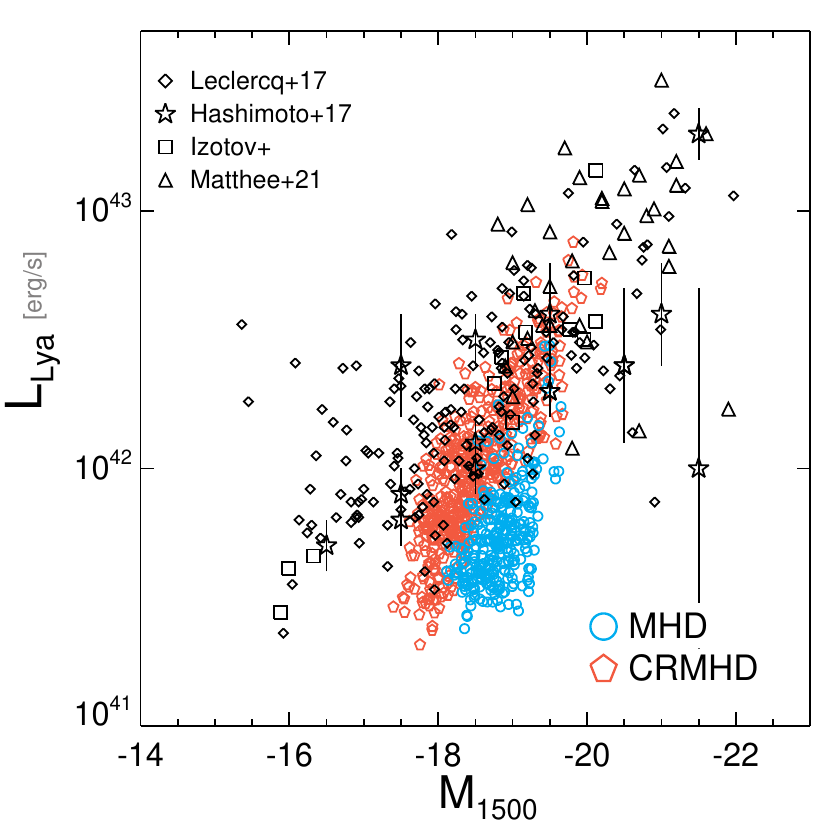}
    \caption{Comparison of the UV magnitudes at 1500\AA\ ($M_{1500}$) and \Lya\ properties of a simulated galaxy with observations. Black symbols indicate the observed samples of LAEs at various redshifts  \citep{leclercq17,hashimoto17,izotov18b,matthee21}, as indicated in the legend. Blue circles and red pentagons indicate simulated UV and \Lya\ from our simulations without and with CR feedback, respectively. Both quantities are dust-attenuated. Each data point represents a different sightline and snapshot between $2 \le z \le 4$. IGM attenuation, which lowers the \Lya\ luminosities by $\sim 15\%$, is not included in this figure. }
    \label{fig:lum}
\end{figure}

In Figure~\ref{fig:lum}, we compare the 1D-projected, attenuated UV magnitude at $1500\, \AA$ ($M_{1500}$) and \Lya\ luminosities with those of the observed LAE samples. Note that here we use the \Lya\ luminosities obtained from the snapshots {\em after} post-processing with {\sc ramses-rt} to make more direct comparisons with observations. Although the \MHD\ galaxies are more actively star-forming than those in the \CRMHD\ run, the attenuated $L_\mathrm{Lya}$ is $\sim$50\% fainter ($\approx3.5\times10^{41}\,\mathrm{erg/s}$ vs. $5.3\times10^{41}\,\mathrm{erg/s}$), as the typical escape fraction of \Lya\ (\fLya) is significantly lower ($\approx0.025$ vs. $0.085$). Similarly, the intrinsic $M_{1500}$ is brighter by $\approx 1$ mag in the \MHD\ run; however, the attenuated $M_{1500}$ is comparable, as the dust attenuation is stronger in the \MHD\ run. As a result, we find that the \CRMHD\ samples are more compatible with the observed LAEs, whereas too little \Lya\ escapes from the \MHD\ sample for a given $M_{1500}$. Part of this vertical offset may be alleviated by resolving the turbulent structures in the hydrodynamic simulations, as these allow a more efficient escape of \Lya\ photons \citep{kakiichi21,kimm19}. However, the ability to disrupt the star-forming clouds early via some collective feedback is more likely to be effective in enhancing the escape fraction, as it directly controls the low-density channels, through which \Lya\, escapes \citep[e.g.,][]{kimm22}. Bearing these differences in mind, we first analyze the simulations without post-processing with {\sc ramses-rt} and compare them with those obtained from the post-processed snapshots\footnote{We discuss the results of the simulations without LyC post-processing, because simulations with CR-driven outflows are often performed without RT. The comparison also provides insights into how observed \Lya\ spectra may be obtained, as discussed later. }.

\subsection{Impact of cosmic rays on Ly$\alpha$ profiles}

\begin{figure}
	\includegraphics[width=\columnwidth]{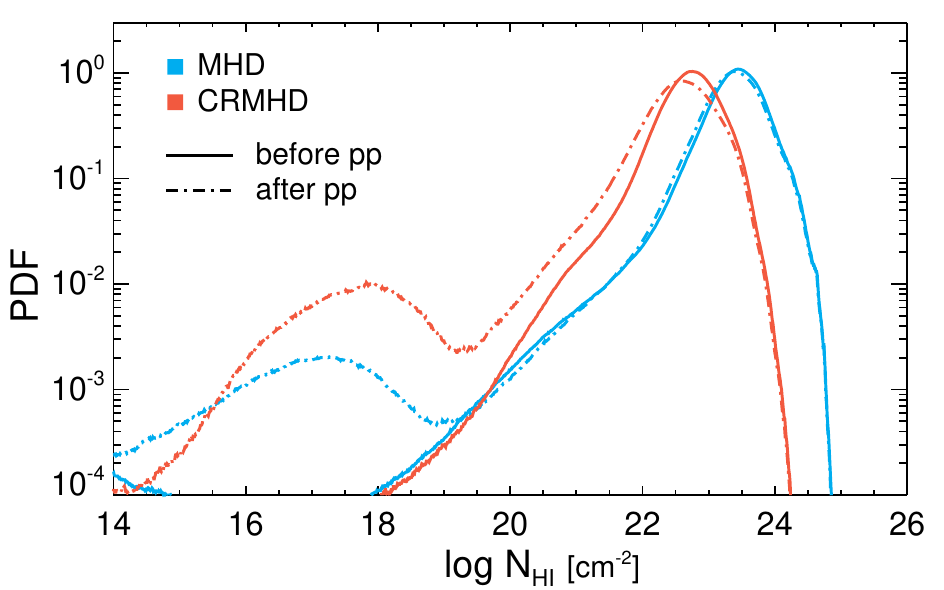}
    \caption{LyC luminosity-weighted distribution of $N_\mathrm{HI}$ measured from each star particle over $2\le z \le 4$. Different color-codes correspond to different simulations, as shown in the legend. Simulations post-processed with {\sc ramses-rt} are shown as dot-dashed lines, while those without post-processing are shown as solid lines.}
    \label{fig:hist_NHI}
\end{figure}

We begin by discussing the impact of CR pressure on HI column density distributions, which are important for \Lya\ radiative transfer. Figure~\ref{fig:hist_NHI} demonstrates that CR feedback not only produces an extended distribution of neutral hydrogen (Figure~\ref{fig:img}) but also lowers $N_\mathrm{HI}$ around young stars. Here, we measure $N_\mathrm{HI}$ from each star particle by casting rays along 3072 sightlines inside the virial sphere using the {\sc Healpix} algorithm \citep{gorski05}. The solid histograms in Figure~\ref{fig:hist_NHI} are obtained by taking LyC luminosity-weighted averages from all the snapshots over the range $2\le z \le 4$ \emph{without} LyC post-processing. Stars in the \MHD\ run are enshrouded by an optically thick gas of $N_\mathrm{HI}\sim 10^{23}\,\mathrm{cm^{-2}}$, whereas the values of  $N_\mathrm{HI}$ in the \CRMHD\ run are smaller by an order of magnitude  (solid lines). A small upturn below $N_\mathrm{HI}\sim 10^{14}\,\mathrm{cm^{-2}}$ is present, but it accounts for less than $0.1\%$ of the total sightlines. As can be seen from Figure~\ref{fig:img}, the reduction in $N_\mathrm{HI}$ due to CR feedback occurs on ISM scales, not CGM scales. When we post-process the simulation output with {\sc ramses-rt} to account for photo-heating by ionizing radiation (dot-dashed lines), the ionized hydrogen fraction is increased, but the decrease in $N_\mathrm{HI}$ is not very significant (see also Figure~\ref{fig:img_pp} for a comparison of $N_\mathrm{HI}$ before and after LyC post-processing). The $N_\mathrm{HI}$ distribution becomes more skewed to lower values after the post-processing, but the luminosity-weighted logarithmic mean shifts only from $\log N_\mathrm{HI}/\mathrm{cm^{-2}}=22.7$ to $22.4$ in the case of \CRMHD\ or from $23.4$ to $23.3$ in \MHD. Thus, the HI distribution remains still more extended, and the inner halo with $r<0.2\, R_\mathrm{vir}$ is filled with $N_\mathrm{HI}\gtrsim10^{17}\,\mathrm{cm^{-2}}$ in \CRMHD, as can be seen in Figure~\ref{fig:img}. 

\begin{figure}
	\includegraphics[width=\columnwidth]{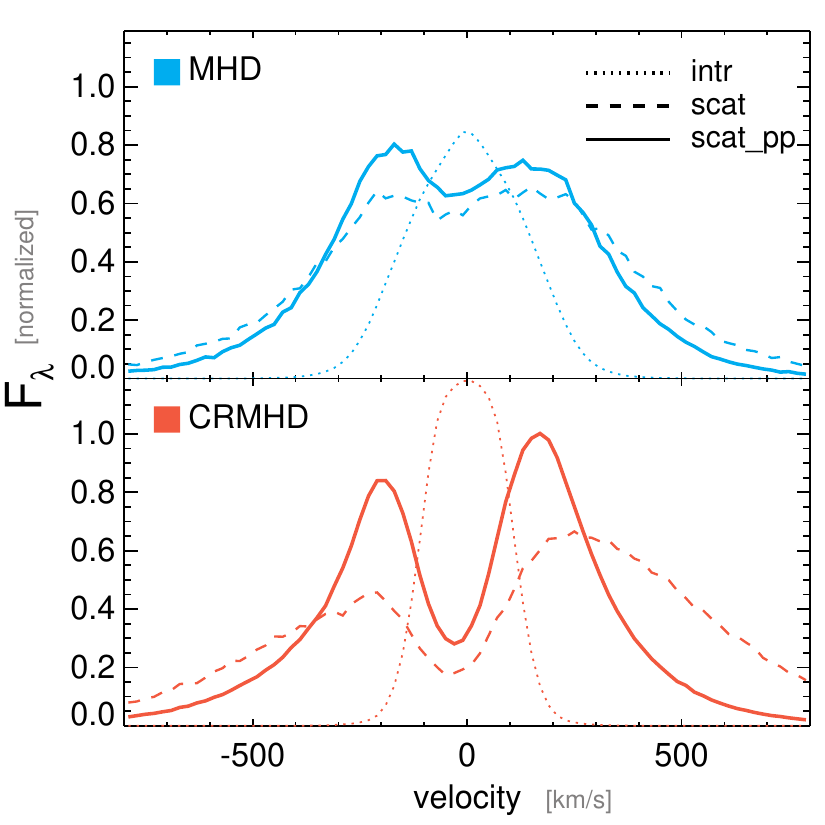}
    \caption{Stacked, angle-averaged \Lya\ profiles from the MHD and CRMHD runs (upper and lower panels, respectively). Profiles are normalized to the maximum flux of the emergent spectrum in the CRMHD run. Intrinsic fluxes are reduced by a factor of two for clarity. The dotted and dashed lines represent the intrinsic  \Lya\ spectrum and results of MCRT calculations on the simulation outputs without LyC post-processing, respectively. The solid lines indicate the MCRT calculations performed on the post-processed simulations, which also include the contribution from collisional radiation. Attenuation by the IGM is not included. }
    \label{fig:spectra}
\end{figure}

\begin{figure*}
\includegraphics[width=0.5\linewidth]{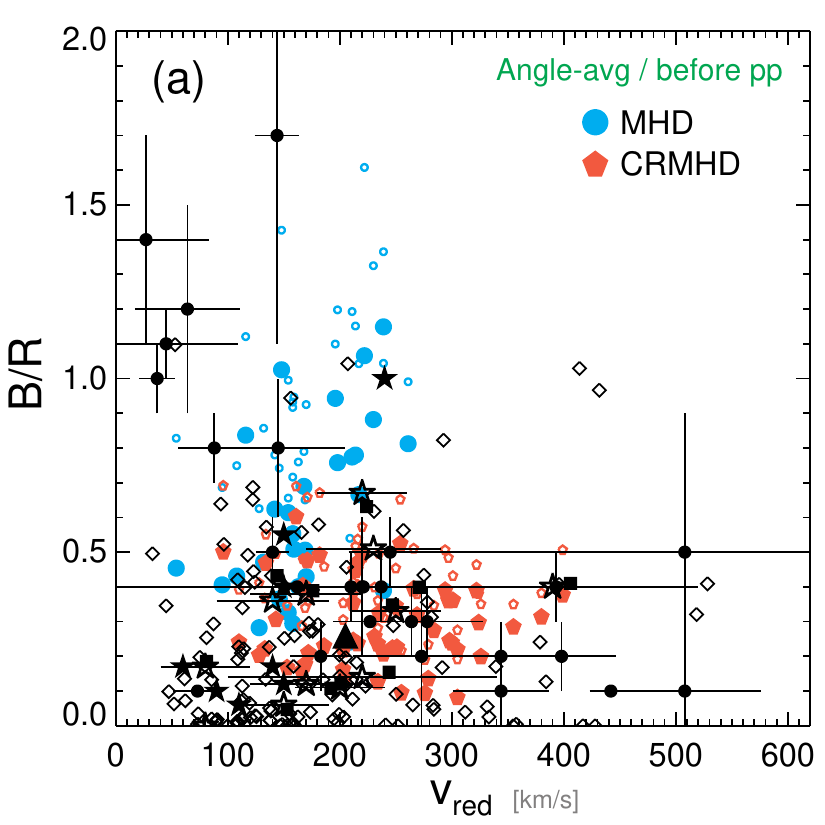}
\includegraphics[width=0.5\linewidth]{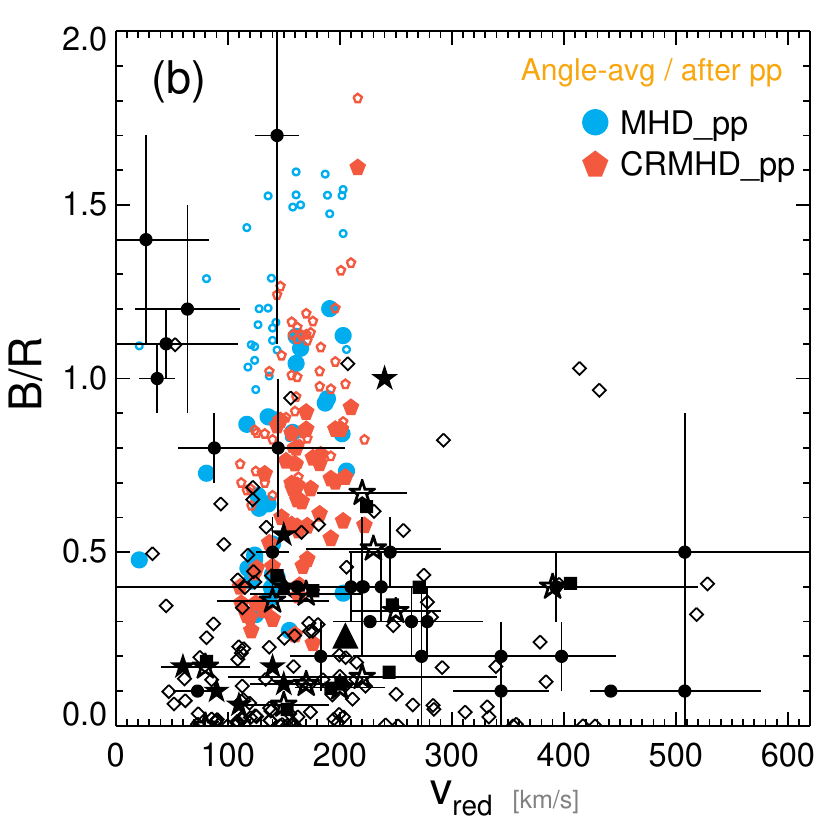}
\includegraphics[width=0.5\linewidth]{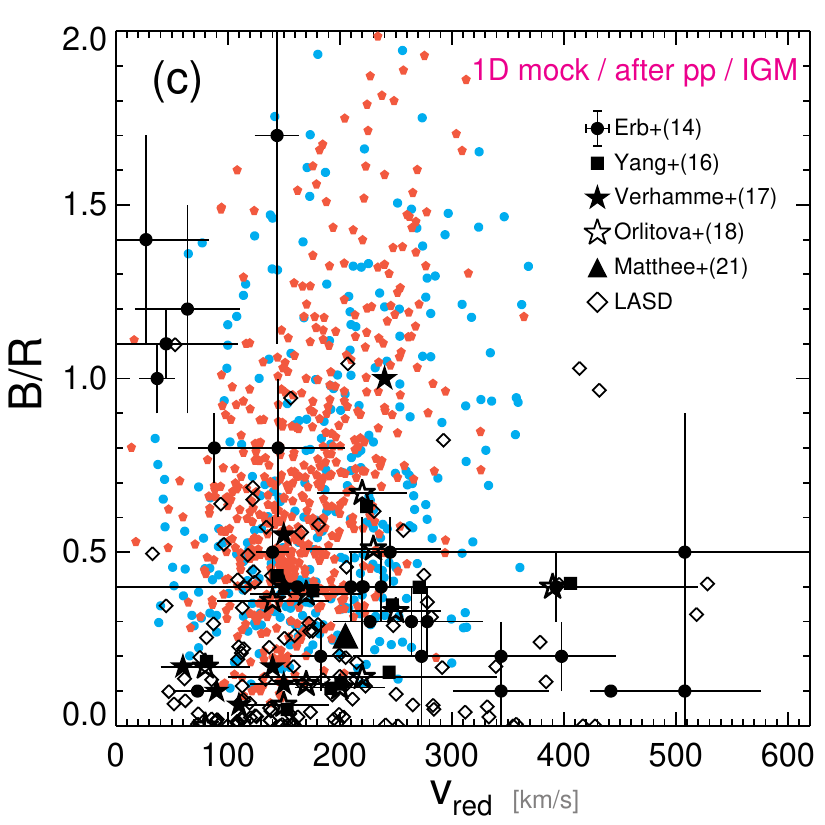}
\includegraphics[width=0.5\linewidth]{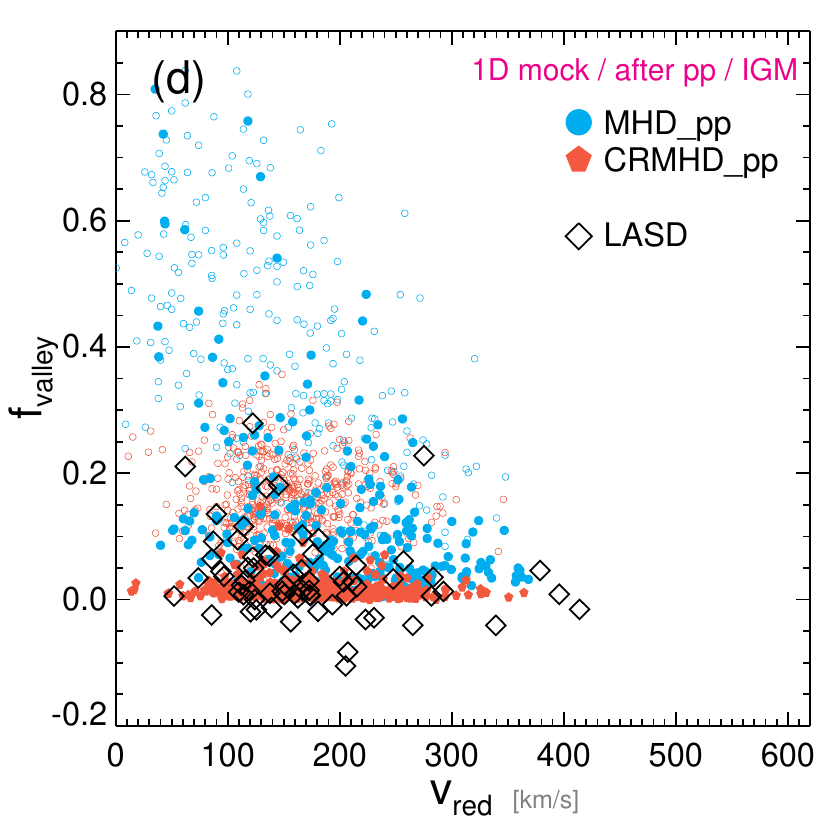}
    \caption{ Comparison of the blue-to-red flux ratio of \Lya\ (\BR) and the location of the red peak (\vred) in velocity space in simulations with (\CRMHD) and without CRs (\MHD). The upper panels show the angle-averaged quantities before (panel-a) and after (panel-b) LyC post-processing with {\sc ramses-rt}. The filled and empty symbols represent measurements after and without applying the simple IGM attenuation model of \citet{Inoue14}, respectively. \Lya\ properties estimated from 1D mock spectra within a 2.5 arcsec aperture are compared in the panel-(c). The black symbols correspond to observational data from \citet{erb14,yang16,verhamme17,orlitova18,matthee21,runnholm21}. Panel-(d) shows the simulated ratio between the flux in the valley of the two peaks and maximum of the peaks (\fvalley,  filled circles). The black diamonds indicate the LASD values obtained using the systematic redshift. For comparison, we also include \fvalley\ values, calculated using \Lya\ collected within the entire virial sphere (empty circles). }
    \label{fig:BR}
\end{figure*}

\subsubsection{Peak locations and blue-to-red flux ratios}

We find that the emergent spectra of the two simulations (before LyC post-processing) are notably different. To focus on the properties of the \Lya\ profiles, we compute the average of the normalized angle-averaged \Lya\ spectrum at $2\lesssim z \lesssim 4$ in Figure~\ref{fig:spectra}.  Although the typical input spectrum (dotted lines) is narrower in \CRMHD, the angle-averaged emergent spectrum (dashed lines) is broader. The flux redward of the \Lya\ spectrum dominates over the blue part ($\BR\sim0.6$), as the galactic outflows are 3--4 times stronger in \CRMHD\, \citep[][see their Figure~ 9]{rodriguez-montero24}. Furthermore, while galactic outflows of the order of a few $\msunyr$ are present at $0.2\, R_\mathrm{vir}$ in the \MHD\ run, $\BR$ is close to 1 because these outflows are mostly ionized and optically thin. In fact, the presence of blue-dominated spectra, even when galactic outflows are present, is not uncommon in other simulations \citep{smith19,smith22,mitchell21,blaizot23}.

The differences become more obvious when individual \Lya\ features are compared. Figure~\ref{fig:BR} (panel-a) shows the location of the red peak in velocity space ($\vred$) and the flux ratio (\BR) from the angle-averaged spectrum of each galaxy at $2<z<4$. Empty symbols correspond to the measurements without IGM attenuation, while the filled symbols show the results after applying the simple IGM model of \citet{Inoue14}. The \CRMHD\ sample shows larger \vred\ and lower \BR\ than \MHD, which is indicative of strong outflows. Additionally, the entire \CRMHD\ sample has red peak-dominant spectra ($\BR<1$) even without IGM attenuation, while 37\,\% of the \MHD\ sample has $\BR>1$. The fraction having $\BR>1$ drops to 10\,\% when IGM attenuation is included. More importantly, some of the \CRMHD\ data exhibit \vred\ up to $\approx400\,\kms$, thus covering the wide range of \vred\ obtained from $z\sim3$ LAEs \citep{erb14} or low-$z$ LAEs taken from the Lyman Alpha Spectral Database (LASD) \citep[][references therein]{runnholm21}. In contrast, the \MHD\ samples show \vred\ of up to $\approx 270\,\kms$, which is significantly less than that shown by \CRMHD, as dust effectively destroys \Lya\ photons before they diffuse in frequency. Taken at face value, the simulation with CR feedback (\CRMHD) appears to predict \Lya\ properties that are more consistent with the LAE observations, while the simulation with the overcooling features (\MHD) can only explain a limited range of galaxies on the \vred-\BR\ plane.

However, we find that more accurate ionization structures make the interpretation non-trivial.  Figure~\ref{fig:BR} (panel-b) shows the \Lya\ properties obtained from the angle-averaged spectra after LyC post-processing with {\sc ramses-rt}. The mean \vred\ in the \CRMHD\ run is remarkably reduced from 234\,\kms\ (left panel) to 162\,\kms (right panel), while the mean IGM-attenuated \BR\ increases from 0.29 to 0.61. Similarly, the \MHD\ run shows a decrease in \vred\ and an increase in \BR, but the changes are less dramatic ($170 \rightarrow 153\,\kms$, $ 0.61\rightarrow 0.69$). As a result, the two different simulations predict similar \Lya\ line features overall, in contrast to the comparison of the simulations before post-processing with {\sc ramses-rt}.  The similar \BR\ values can be partly attributed to IGM absorption, which suppresses the blue flux more strongly in the \MHD\ sample.

To understand why \vred\ is significantly reduced in the \CRMHD\ run, we examine a snapshot at $z=2.7$, in which the galaxy exhibits the largest \vred\ ($\approx 400\,\kms$), but a smaller $\vred$ of $\approx 170\,\kms$ after post-processing. This galaxy undergoes a minor burst phase with a star formation rate of $3\,\msunyr$; however, it is not associated with mergers. The effective HI column densities ($N_\mathrm{HI,eff}$) measured from the LyC escape fraction are very similar before ($1.6\times10^{18}\,\mathrm{cm^{-2}}$) and after post-processing ($9.8\times10^{17}\,\mathrm{cm^{-2}}$), but the \Lya\ escape fraction increases by an order of magnitude (1.1\% vs 10.3\%). Here, we use  $N_\mathrm{HI,eff}=-\log_{10} f_\mathrm{LyC}^\mathrm{noDust}/\sigma_\mathrm{LyC}$, where $f_\mathrm{LyC}^\mathrm{noDust}$ is the luminosity-weighted escape fraction of the LyC photons, measured from each star particle within the virial sphere in the absence of dust, and $\sigma_\mathrm{LyC}=6.3\times10^{-18}\,\mathrm{cm^{-2}}$ is the absorption cross-section at the Lyman edge. As we assume that LyC and \Lya\ photons are produced at the same positions, the marked difference demonstrates that \Lya\ is very efficient in finding and propagating along low-density channels, thus making it less sensitive to the mean optical depth of the volume-filling gas inside the virial sphere. Figure~\ref{fig:hist_NHI}  supports this interpretation by showing that the fraction of sightlines with  $N_\mathrm{HI} \le 10^{18}\,\mathrm{cm^{-2}}$ increases from $\le 0.1\%$ to only $\sim1\%$ (before and after LyC post-processing), but this change is sufficient to alter the \Lya\ profiles.

Ly$\alpha$ is highly sensitive to the geometry and velocity structures of the scattering medium, leading to significant variations in spectra along different sightlines compared with angle-averaged spectra \citep[e.g.,][]{blaizot23}. The effect of orientation can be seen in Figure~\ref{fig:BR} (panel-c), which shows the line properties computed from the IGM-attenuated, 1D mock spectra. To make a fair comparison with existing observed profiles, such as those from the HST Cosmic Origins Spectrograph, we include a collisional term and collect \Lya\ within a 2.5 arcsec aperture.\footnote{We confirm that collecting \Lya\ within the entire virial radius does not significantly affect \BR. However, the number of sightlines showing large \vred\ ($\gtrsim 300\,\kms$) increases notably in \MHD\ ($7\rightarrow24$), as the contribution from \Lya\ photons with smaller velocity shifts in the outer halo is excluded \citep[see e.g.,][]{guo24}.} For both simulations the distributions of \vred\ and \BR\ along the various sightlines are broader, with $\BR \lesssim 2$ and $\vred \lesssim 350\,\kms$, in contrast with the angle-averaged values. The remarkable similarity in \vred\ and \BR\ distributions predicted by the two simulations, with and without CR-driven winds,  suggests that these two metrics might not be enough to distinguish different feedback models, unless very large observational samples are available.

Nevertheless, the comparison with observations provides some useful insights into gas kinematics.  The \CRMHD\ galaxy shows a median \vred\ of 151\,\kms, comparable to that of the local \citep{yang16,verhamme17,orlitova18,runnholm21} or high-$z$ LAEs \citep{erb14}. The IGM-attenuated \BR\ (0.64, median) is also broadly consistent with the observations, but tends to be higher than the average value of the \citet{erb14} sample (0.4) or local LAE values of $\approx 0.1$ \citep{runnholm21}. Without the IGM attenuation, the simulated \BR\ increases to $\approx 1.0$. Moreover, a significant fraction (21\%) of the simulated profiles with IGM attenuation are blue-dominated ($\BR>1$), while the fraction appears lower ($\sim 13\%$) in a sample of 32 LAEs at $z \sim 2.7$ \citep{trainor15}. Although these blue-dominated galaxies could possibly represent quiescent phases and are thus more difficult to observe \citep{blaizot23}, the overprediction of \BR\ indicates that both simulations may be lacking outflows. The paucity of galaxies with large \vred\ ($\gtrsim 350\,\kms$) is likely a related problem, although it is equally possible that the large \vred\ values in the observations arise simply because the observed samples are brighter than the simulated ones, as discussed further in Section~\ref{sec:dis_sub1}.

\subsubsection{Valley flux}

\citet{gronke18} showed that the impact of CR-driven winds is more evident when comparing the flux ratio (\fvalley) between the valley ($F_\mathrm{valley}$) and maximum ($F_\mathrm{max}$) of the two peaks. Their ISM simulations without CRs produce \Lya\ spectra with a high value of $0.3\la \fvalley \la 0.7$, while \fvalley\ in their CR simulations can be as small as $\sim 0.1$ owing to the suppression of \Lya\ escape at the line center by  the extended neutral outflows. We also measure the flux ratio from 1D mock spectra extracted within a 2.5 arcsec aperture and compare it with the recent LASD data from \citet{runnholm21} in Figure~\ref{fig:BR} (panel-d). Because measuring the valley flux requires a high spectral resolution and can therefore be uncertain depending on the instrument used, we include only those samples for which $F_\mathrm{valley}$ measurements from two independent redshift determination methods are available and agree within $\Delta F_\mathrm{valley} < 0.2$ for conservative estimates.

We find that simulations without CRs predict a mean valley flux ratio of $\left<\fvalley\right>=0.12\pm0.13$, while \fvalley\ values are further reduced in the presence of CR-driven winds ($0.02\pm0.02$), where the range represents the standard deviation. The sightlines from the \CRMHD\ run show greater  overlap with the the observed distribution of $\left<\fvalley\right>\approx0.04\pm0.07$, while a non-negligible fraction of the \MHD\ run exhibits large \fvalley\ values of $\gtrsim 0.2$. However, both simulations yield results that are broadly consistent with observations. This level of agreement is somewhat unexpected, particularly in the \MHD\ run, considering that its angle-averaged profile shows a prominent flux at the line center in Figure~\ref{fig:spectra}. 

To illustrate the importance of aperture effects in interpreting the \Lya\ profiles, we remeasure \fvalley\ by collecting \Lya\ within the entire virial sphere (shown as empty circles in Figure~\ref{fig:BR}, panel-d). In contrast to \fvalley\ measured within the central $\sim 10\,\mathrm{kpc}$ radius (i.e., 2.5 arcsec diameter), the flux at line center in the \MHD\ run becomes much more prominent ($\left<\fvalley\right>=0.40\pm0.19$), indicating that \Lya\ from the outer halo contributes preferentially near line center \citep[e.g.,][]{Erb2023,guo24}. The \fvalley\ values within the virial sphere from the \CRMHD\ run are also increased to $\left<\fvalley\right>=0.17\pm0.06$, but they remain significantly lower than those in \MHD. This difference can be attributed to the fact that, in the \MHD\ simulation, $\approx 97\%$ of the \Lya\ photons are destroyed by dust owing to high column densities, thus making the emergent signal more sensitive to photons originating in low-density regions. By directly comparing the emergent line profiles between \MHD\ and \CRMHD\ runs, we confirm that the valley flux is not necessarily high in the \MHD\ run, but the flux at reasonably large velocity shifts, and thus $F_\mathrm{max}$, is significantly suppressed.

Our findings support the conclusion that CR-driven winds can lead to lower \fvalley\ values \citep[cf.][for ISM patch simulations]{gronke18}. However, low \fvalley\ can also result from a small observational aperture, which preferentially captures photons that undergo multiple scatterings through an optically thick ISM. In this context, the behavior of \fvalley\ measured over larger spatial scales may be more useful, as it may reflect the angle-averaged properties of the \Lya\ spectra better, thereby motivating further investigation.

\subsection{Ly$\alpha$ surface brightness profiles}

The previous section showed that the presence of ionized channels allows efficient escape of \Lya. These escaped photons will subsequently scatter with the neutral HI in the CGM. An enhanced HI spatial distribution, resulting from CR feedback, could enhance this scattering and potentially leave an imprint on the surface brightness (SB) profiles. To investigate this possibility, we median-stack the SB profiles obtained along 10 different sightlines per snapshot and combine the results for two different redshift ranges $2\le z \le 3$ and $3\le z \le 4$. Following \citet{mitchell21}, the SB profiles are PSF-convolved using a Moffat function with $\beta=2.8$ and $\mathrm{FWHM/''}=0.875-(\lambda/10^4\,\AA)/3$. To correct for the redshift dimming effect, we multiply the brightness by $(1+z)^4$ and assume that all the samples are placed at $z=3$. Again we use the simulation data, post-processed with ionizing radiative transfer for both \CRMHD\ and \MHD\ runs, and include the contribution from collisional radiation.  The intrinsic \Lya\ luminosity from collisional radiation in the \CRMHD\ run is, on average, $\sim3$ times brighter than that in \MHD, but it accounts for only $\approx 15\%$ of the total \Lya.

\begin{figure}
	\includegraphics[width=\columnwidth]{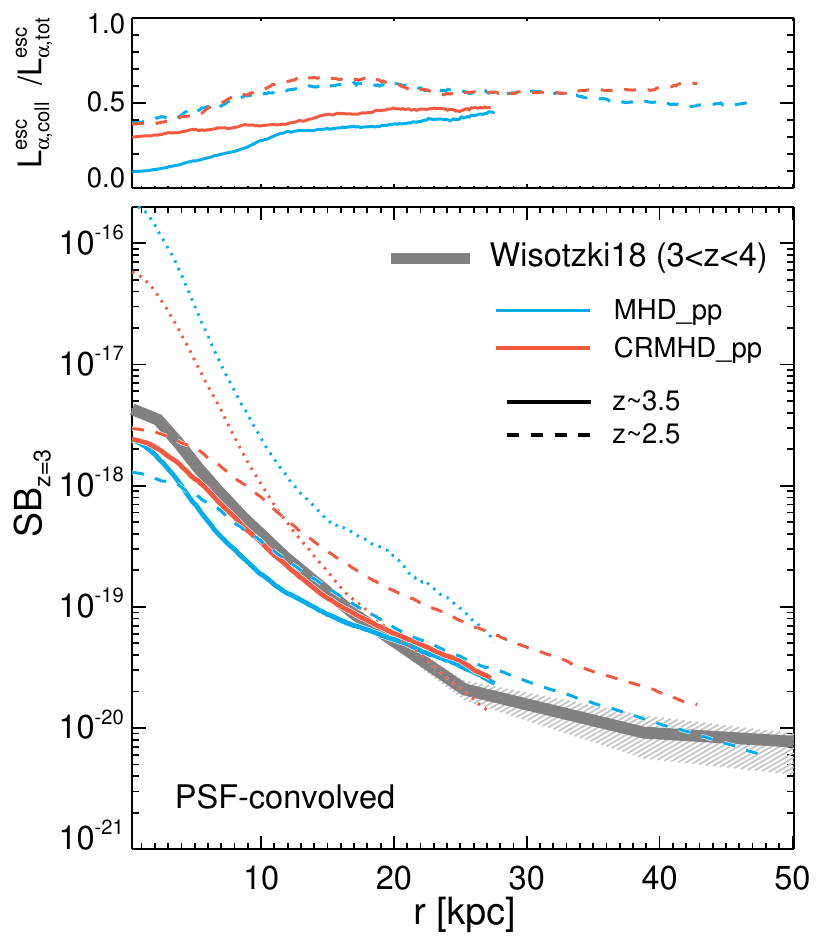}
    \caption{ Upper panel: fractions of collisional radiation to the total emergent, PSF-convolved \Lya\ SB profiles. Solid and dashed lines show results for different redshift ranges ($2<z<2.5$ and $3<z<3.5$, respectively). Lower panel: median-stacked SB profiles of galaxies simulated without (\MHD, cyan) and with (\CRMHD, orange) CR feedback.  The median-stacked SB profile of LAEs at $z\sim3.5$, obtained with the MUSE instrument, is included for comparison  \citep[thick gray line with shaded regions,][]{wisotzki18}. We also include the PSF-smoothed intrinsic SB profiles as dotted lines. To correct for the redshift dimming effect, the SB profiles are scaled by a factor of $(1+z)^4/(1+3)^4$. }
    \label{fig:sbp}
\end{figure}

Figure~\ref{fig:sbp} (lower panel) shows that \Lya\ emission is more extended in the CR feedback run for a given observation threshold. As aforementioned, although the intrinsic \Lya\ luminosity of the \CRMHD\ sample is half as bright as that of \MHD, \Lya\ escapes more efficiently. As a result, the attenuated luminosities are, on average, 1.9 times brighter in \CRMHD\  than in \MHD. This is evident when comparing the colored solid lines, where the thin lines indicate the PSF-smoothed total intrinsic SB profiles (including collisional radiation), while the thick lines denote the PSF-smoothed emergent profiles at $3<z<4$. In particular, the emergent SB profile from the \CRMHD\ run at $r>20\,\mathrm{kpc}$ is brighter than the intrinsic profile, clearly indicating that scattering with the neutral CGM  facilitates more extended \Lya\ emission.   Interestingly, in the case of the \MHD\ run, the inefficient escape of \Lya\ is seen not only in the central region, but also in the outer halo. This is due to the fact that satellite galaxies in the \MHD\ run are overcooled and contribute a significant fraction of intrinsic radiation even at $10\lesssim r \lesssim 30\,\mathrm{kpc}$. 

It is also interesting to compare \Lya\ profiles originating from collisional radiation. Because cooling radiation is produced in more extended regions, it escapes more easily ($f_\mathrm{Lya,coll}\sim0.5$) than that from young stars and is therefore more affected by the CGM. For example, the effective radius of the intrinsic cooling radiation in \CRMHD\ is three times larger ($7\,\mathrm{kpc}$ before PSF smoothing) than that of the young stars ($2.3\,\mathrm{kpc}$). Although not shown, we find again that at $r\gtrsim20\,\mathrm{kpc}$, the emergent cooling radiation in the \CRMHD\ run becomes brighter than the intrinsic emission, while the emergent SB profile in \MHD\ remains fainter than its intrinsic profile at all radii; this supports our interpretation of the impact of CR driven outflows on the extended \Lya\ SB profile.

The SB profiles simulated with CR feedback are in good agreement with those of the observed LAEs.  The gray thick line in Figure~\ref{fig:sbp} illustrates the median-stacked SB profile of 92 LAEs obtained with MUSE at $3 < z < 4$ \citep{wisotzki18}. Note that the MUSE sample has UV magnitudes comparable to those of the \CRMHD\ galaxies in the same redshift range. Our simulations at $z\sim3.5$ predict \Lya\ profiles that closely match the observations, although the SB from the \MHD\ run is slightly fainter. As expected, the simulations at lower redshifts ($z\sim2.5$) yield brighter profiles than their $z\sim3.5$ progenitors, as star formation becomes more active in more massive galaxies.  The more luminous LAE subsample  in \citet{wisotzki18} with $>10^{42}\,\mathrm{erg\,s^{-1}}$ indeed shows systemically brighter SB profiles \citep[see also][]{kusakabe22}.

Using a cosmological radiation-hydrodynamic simulation of a dark matter halo of mass $10^{11}\,\msun$ at $z=3$,  \citet{mitchell21} successfully reproduce the SB profile of \citet{wisotzki18} and argue that collisional excitation alone accounts for half of the SB in the CGM region ($r\gtrsim10\,\mathrm{kpc}$). Figure~\ref{fig:sbp} (upper panel) shows that in our simulations, both with and without CR pressure, collisional radiation also explains $\sim30$--$50\%$ of the total \Lya\ SB in the CGM region \citep[see also][]{smith19}. However, our simulations reveal additional complexity in accounting for the remaining SB. In the \CRMHD\ run, the CR-driven outflows shape the extended \Lya\ halo by reducing the surrounding gas density near the young stars. This, in turn, facilitates the escape of more photons and scattering in the CGM, resulting in a more extended \Lya\ halo. In contrast, in the over-cooled \MHD\ galaxy, a large amount of \Lya\ is produced and destroyed, with only a small fraction reaching the CGM.  Although CR feedback is not included in \citet{mitchell21}, enhanced \Lya\ escape is achieved via boosted SN feedback. Based on the results from the IllustrisTNG simulations \citep{pillepich2018,nelson2019_tng50}, \citet{byrohl21} show that  the SB profiles from \citet{leclercq17} can be reproduced if dust is absent in low-mass galaxies. All of these possibilities suggest that no unique mechanism exists for generating extended \Lya\ SB, and multiple scenarios must be explored in parallel to constrain galaxy formation models using SB observations.

\section{Discussion}
\label{sec:discussion}

\subsection{Lack of \Lya\ line profiles with strong outflow features}
\label{sec:dis_sub1}

The previous section showed that a majority of galaxies post-processed with {\sc ramses-rt}  have \Lya\ features consistent with those of local star-forming galaxies \citep[e.g.,][]{yang16,verhamme17,orlitova18}. However, only a few simulated samples have a \Lya\ profile with both a large \vred\ of $\gtrsim 350\,\kms$ and a low $\BR$  of $\lesssim0.5$. These combined features can be interpreted as indicative of strong neutral outflows \citep[e.g.,][]{verhamme06,dijkstra06}, thus motivating further discussion of the comparison between simulations and observations. 

To understand the origin of a large \vred, we first focus on the \CRMHD\ galaxies that show $\vred\sim 400\,\kms$ in the spectra computed without LyC post-processing (see panel-a in Figure~\ref{fig:BR}). To determine whether the ISM or CGM is primarily responsible for shaping the \Lya\ profile, we perform MC \Lya\ RT twice, once without the ISM and once without the CGM. For simplicity, collisional radiation is neglected in this exercise. Separating the ISM from the CGM is non-trivial, and we simply assume that the gas within $0.1\,\rvir$ ($\approx5\,\mathrm{kpc}$) is more associated with the ISM. This is motivated by the fact that $R_\mathrm{90}$  is $\approx 4.5\,\mathrm{kpc}$ at this redshift, where $R_\mathrm{90}$ is the radius within which 90\% of the UV light is confined. Note that the \Lya\ source positions are taken from stellar positions within $r<0.1\,\rvir$ and thus differ slightly from the source positions used in Figure~\ref{fig:BR}, where all the stellar particles within $r<\rvir$ are used along with collisional radiation. In the case of the sightline with the largest \vred\ of $400\,\kms$, only $\sim1\%$ of the \Lya\ photons escape their virial halo and show \vred\ of $\sim 250\,\kms$ when \Lya\ photons interact only at $r<0.1\,\rvir$\footnote{Unlike the other cases, this particular snapshot has a strong \Lya\ emission from satellites, which complicates the interpretation of the profile in terms of $N_\mathrm{HI}$; however, the ISM region still plays a more important role than the CGM in setting the large \vred.}. Conversely, if the scattering occurs only in the outer region ($r>0.1\,\rvir$), $\sim 97\%$ of them leave the halo with $ \vred\sim150\,\kms$. In the latter case, the spectrum is dominated by the blue peak ($\BR=1.6$ without IGM attenuation), indicating that a significant fraction of the neutral gas is collapsing in the outer region \citep{guo24,yuan24}. Indeed, we confirm that the volume-filling neutral hydrogen in this simulated galaxy tends to be inflowing rather than outflowing.

\begin{figure}
\includegraphics[width=\columnwidth]{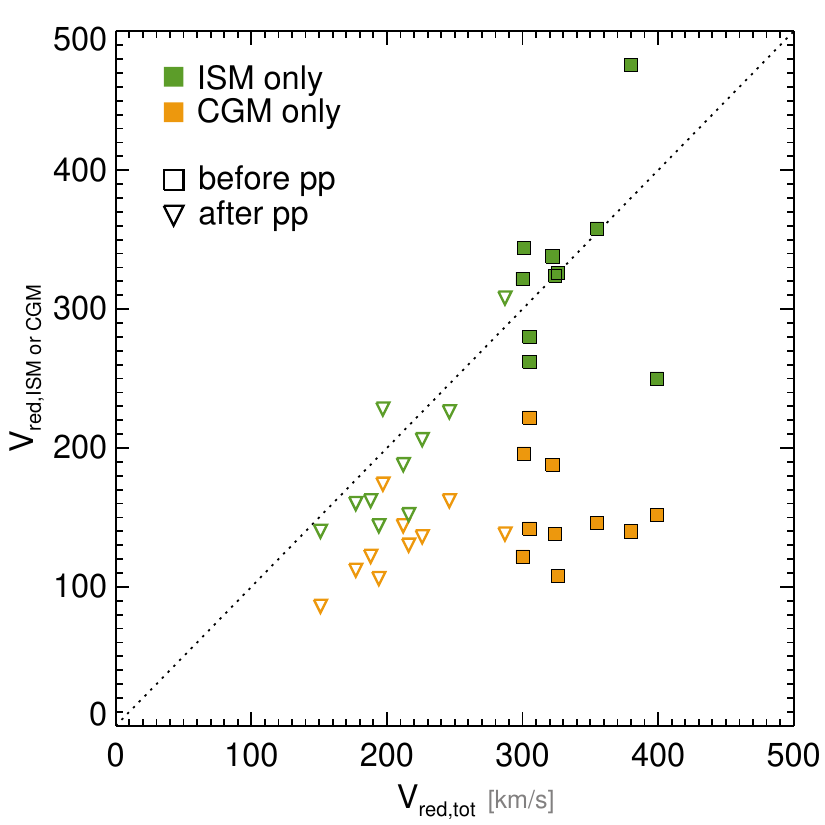}
    \caption{Predicted \Lya\ profiles formed by ISM region ($r<0.1\,\rvir$, green symbols) or by CGM ($r>0.1\,\rvir$, orange symbols). Note that the source positions are limited to $r<0.1\,\rvir$ for $v_\mathrm{red,ISM}$ or $v_\mathrm{red,CGM}$ for easier interpretation. The $v_\mathrm{red,tot}$ values are the same as the \vred\ values in Figure~\ref{fig:BR} (left panel) and thus are obtained from all the photons generated within the virial sphere. The location of the red peak from the galaxy (\vred) is compared before (squares) and after (triangles) LyC post-processing with \sc{ramses-rt}.}
    \label{fig:large_vred}
\end{figure}

To be more quantitative, we repeat this exercise for all snapshots showing $\vred\ge300\,\kms$ in their angle-averaged spectrum in the \CRMHD\ run before LyC post-processing. Figure~\ref{fig:large_vred} (filled symbols) shows that although the CGM alone can produce \vred\ of 100--200 \,\kms, the peak location is primarily determined by the gas within $0.1\,\rvir$. If the scattering medium were uniform with a fixed temperature of $10^4\,\mathrm{K}$ (Equation~\ref{eq:vred}), the \vred\ values inside and outside $0.1\,\rvir$ would imply that the scattering $N_\mathrm{HI}$ in the inner region is an order of magnitude larger than that of the CGM ($\sim6\times10^{20}$ vs. $7\times10^{19}\,\mathrm{cm^{-2}}$).  \Lya\ processed by the inner gas often has larger \vred\ values compared with those influenced by the ISM and CGM together (\vred\ in the left panel of Figure~\ref{fig:BR}). However, this difference is simply due to the different source positions within $0.1\,\rvir$, as shown in Figure~\ref{fig:large_vred}, which was intended to minimize the complexity of the source geometry due to satellites. 

The contribution of the inner gas to \vred\ in the \CRMHD\ run decreases when more precise ionization structures are used (open symbols), but it remains the dominant component shaping the line profile compared with the CGM. Comparing the relative shift between the filled and empty symbols in Figure~\ref{fig:large_vred}, the decrease in \vred\ can be attributed to the change in the ionization structures of the inner halo rather than that of the CGM. We recall that $N_\mathrm{HI}$ around young stellar populations does not vary significantly before and after the LyC post-processing, suggesting that this subtle change in the ionized structure of the ISM region leads to the significant decrease in \vred. In fact, the analytical calculation (Equation~\ref{eq:vred}) suggests that $N_\mathrm{HI}$ needs to be of the order of $10^{21}\,\mathrm{cm^{-2}}$ to accommodate the large $\vred$ of $\gtrsim 400$--$500\,\kms$, but the actual $\vred$ is significantly smaller even though the column density distributions in our simulations peak around $N_\mathrm{HI}\sim10^{22}\,\mathrm{cm^{-2}}$ (Figure~\ref{fig:hist_NHI}) \citep[see also][]{gronke16a}.

High-resolution galactic-scale simulations show that the ISM is highly turbulent and mostly supported by warm structures owing to the self-regulation of star formation processes \citep[e.g., ][to name a few]{hennebelle14,gatto17,kimcg18}. Such turbulent structures seem to be missing in the simulations presented here. This may be due to finite resolution, the absence of radiation feedback \citep{rosdahl15shine,grudic18,kimjg18}, a simple chemical network \citep{katz22prism,kimjg23}, or an inefficient star formation model \citep{agertz15,kang25}. Because even a small fraction of sightlines with low $N_\mathrm{HI}$ can significantly reduce the \Lya\ line width, the development of the multiphase ISM would help \Lya\ photons escape more freely, thereby further reducing the number of scatterings with dusty cold ISM and increasing $L_\mathrm{Lya}$. On the other hand, heating the cold dense phase into the volume-filling warm neutral phase by effective stellar feedback could populate relatively high $N_\mathrm{HI}$ lines of sight with little dust, potentially enhancing \vred. Thus, further investigations are needed to determine whether the structural properties of the warm ISM actually facilitate enhanced \Lya\ scattering. An example based on the TIGRESS ISM model \citep{kimcg18} shows the angle-averaged \Lya\ spectra with $\vred\sim 200\,\kms$ \citep{seon20}, which may not be sufficient to explain the observed \vred, even though galactic outflows extending into the CGM could potentially lead to higher \vred\ values.

Another missing component of simulated galaxies is the multiphase structure of the diffuse, low-density CGM. Increased resolution in the CGM helps to better resolve thermally unstable gas and hydrodynamic instabilities, leading to the development of more cool structures \citep{hummels19,peeples19,suresh19,rey24}. This, in turn, could enhance the cooling radiation in the CGM \citep[e.g.,][]{haiman00}. A similar idea has been proposed by \citet{cantalupo14}, who claim that substantial clumping is required to explain observations of giant Lyman alpha nebulae.  Considering that the stellar mass of the \CRMHD\ galaxy lies slightly above the average stellar mass-to-halo mass relation \citep[e.g.][]{behroozi13}, more efficient feedback may still be required, and cooling radiation may be enhanced owing to the increased gas content in the CGM. However, a highly structured CGM is unlikely to boost \Lya\ scattering, making the detection of galaxies with large \vred\ difficult. 

Admittedly, direct comparisons between observations and simulations should be made with caution. Some observed samples with large \vred\ are detected with low signal-to-noise ratio at finite spectral resolution, and the precise determination of \vred\ and \BR\ faces uncertainties, as indicated by the large errors in Figure~\ref{fig:BR}. Additionally, we simulated only one galaxy, whereas observational data likely correspond to different phases of galaxy evolution \citep[c.f. see][for an example that recovers diverse \Lya\ profiles using a single galaxy]{blaizot23}. Increasing the number of mock spectra could also help explain rare features, as the dependence of line properties on sightline variations can be significant \citep{blaizot23,yuan24}. Indeed, although the angle-averaged \vred\ is small ($\la 250\,\kms$), the \CRMHD\ run reveals a few cases with a large $\vred$ of $\sim350\,\kms$ (Figure~\ref{fig:BR}); however, the fraction of galaxies with such large \vred\ values appears too small to be consistent with observations. This may be partly because galaxies with large \vred\ in \citet{erb14} tend to be brighter ($-21 \lesssim M_\mathrm{UV} \lesssim -19$) than our simulated galaxies. \citet{trainor15}, for example, report that galaxies with $\vred\gtrsim400\,\kms$ are more commonly observed at $M_\mathrm{UV}\lesssim -20$. Therefore, future studies are needed to simulate a brighter sample to make a more rigorous comparison and draw a firm conclusion on the paucity of galaxies with large \vred.

% Example figure
\begin{figure}
	\includegraphics[width=\columnwidth]{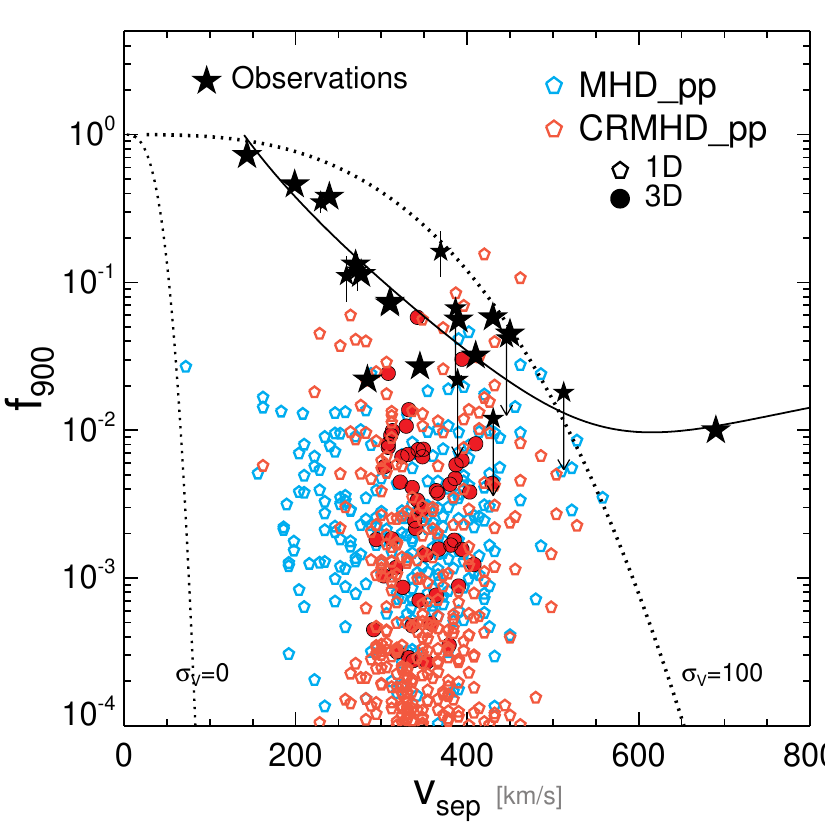}
    \caption{Correlation between separation of double peaks in \Lya\ (\vsep) and escape fraction at $900\,\AA$ (\fLyC). The black star symbols represent the observed properties of LAE galaxies from \citet{verhamme17}, while the colored points denote the results from simulations (empty: 1D, filled: 3D). The black solid line is the empirical relation from \citet{izotov18c}. For comparison, we include the analytic correlation between \vsep\ and \fLyC\ for dust-free uniform media with turbulence velocities of 0 and 100 \kms\ (dotted lines). No clear correlation is found between \fLyC\ and \vsep\ at $\fLyC\la0.1$. }
    \label{fig:vsep}
\end{figure}

\subsection{Implications of sensitive behavior of \Lya}

Using two idealized configurations of homogeneous or clumpy spherical shells, \citet{verhamme15,verhamme17} suggested that the separation of the blue and red peaks (\vsep) could be used to select the LyC leakers. The idea is that galaxies with low-density channels that allow LyC photons to escape would experience fewer scatterings, resulting in a low $\vsep$ of $\lesssim 300\,\kms$.  Based on the Cosmic Origin Spectrograph data of five local Lyman continuum leakers and results from the literature, \citet{izotov18b,izotov18c} further argue for a tight correlation between \vsep\ and the LyC escape fraction \citep[see also][]{flury22b,naidu22}, thus supporting the claim by \citet{verhamme15}. 

In Figure~\ref{fig:vsep}, we compare \vsep\ with the LyC escape fraction at $900\,\AA$ (\fLyC).  Observational data from \citet{verhamme17,izotov18a,izotov18b,izotov21} are shown as star symbols, along with an empirical relation by \citet{izotov18c}. Orange empty hexagons correspond to measurements along different sightlines in the \CRMHD\ run at different redshifts, while cyan hexagons represent measurements from the \MHD\ run. As already shown in Figure~\ref{fig:BR}, the two simulation sets do not exhibit a clear difference in \vsep. However, the \CRMHD\ sample tends to display lower \fLyC\ than does the \MHD\ run, as the volume-filling, extended HI in the CR run absorbs LyC radiation more efficiently \citep[see][for results from cosmological simulations]{farcy25}. Nevertheless, distinguishing between the two models in the \vsep-\fLyC\ plane remains non-trivial owing to the significant overlap.

Our results further suggest that the \vsep-\fLyC\ correlation is likely weak among non-LyC leakers, regardless of the model assumptions. Although some simulated samples overlap with the properties of local galaxies \citep[e.g.,][]{verhamme17},  virtually no correlation exists between \vsep\ and \fLyC\ at $\fLyC \lesssim 0.1$. This is true even when comparing angle-averaged quantities (red filled circles), suggesting that the highly sensitive nature of \Lya\ prevents one from inferring the escape fraction at these low \fLyC\ regimes. Using a large number of simulated galaxies at $z\gtrsim5$, \citet{choustikov24} also reach the same conclusion, while the majority of LyC leakers still exhibit $\vsep \lesssim 300\,\kms$. The considerable scatter found at $\fLyC \lesssim 0.1$ in recent observations \citep{flury22b,naidu22} also appears consistent with our interpretation.

Another intruiguing question is whether the \Lya\ peak location can be used to trace the outflow velocity of the neutral medium. Using a shell model, \citet{gronke15} demonstrate that the two most important parameters, $N_\mathrm{HI}$ and outflow velocity, can be reasonably recovered by fitting the \Lya\ profiles. However, extracting the information becomes progressively difficult when the medium is highly clumpy \citep[e.g.,][]{gronke16a}. Our results also support this, showing that the presence of low-density channels can significantly alter the emergent  \Lya\ spectra. Moreover, sightline-to-sightline variations in \Lya\ profiles are substantial.  A recent study by \citet{yuan24} also showed that \vsep\ is not strongly correlated with HI-weighted velocity, although sightlines with low \BR\ generally trace outflows. Indeed, analyzing the UV and \Lya\ spectra of local star-forming galaxies from the Lyman Alpha Reference Sample, \citet{rivera-thorsen15} find little correlation between \vred\ and the outflow velocity inferred from other low-ionization metal absorption lines. Given these complexities, extracting physical properties from individual spectra may be challenging. Instead, simultaneously reproducing SB profiles and \Lya\ line features for a statistical sample may be more useful to infer general kinematic properties of the neutral medium and constrain feedback models in simulations \citep[e.g.,][]{song20}.

\section{Summary and Conclusions}
\label{sec:conclusions}

This study investigated the impact of CR feedback on the \Lya\ emission properties of a simulated galaxy embedded in a $10^{11}\,\msun$ dark matter halo at $z=2$.  To this end, we used cosmological MHD simulations with (\CRMHD) and without CRs (\MHD)   \citep{rodriguez-montero24} and post-processed them with ionizing radiation transfer using {\sc ramses-rt} to obtain more accurate hydrogen ionization structures. We  then performed Monte Carlo \Lya\ radiative transfer with {\sc rascas} \citep{michel-dansac20} and measured several line properties: the relative peak strength (\BR), location of the red peak (\vred), and relative flux ratio between the valley and the maximum of the two peaks (\fvalley). The comparison of simulated \Lya\ with and without CRs, together with the observed \Lya\ features of LAEs, revealed the following key features.

\begin{itemize}

\item The simulation with CR-driven outflows reasonably reproduces the UV and \Lya\ luminosities of observed LAEs. In contrast, the simulation without CRs predicts galaxies that are fainter in \Lya\ for a given $M_\mathrm{1500}$, compared with the observations. (Figure~\ref{fig:lum}). Although galaxies simulated with CRs form fewer stars by a factor two, their attenuated \Lya\ luminosities are two times brighter, as \Lya\ escapes more efficiently owing to a lower $N_\mathrm{HI}$ distribution (Figure~\ref{fig:hist_NHI}). 

\item Despite the significant difference in $N_\mathrm{HI}$, the resulting \vred\ computed from angle-averaged line profiles are very similar between the \CRMHD\ and \MHD\ runs (Figure~\ref{fig:BR}-b).  Optically thick, dusty galaxies from the \MHD\ run predict an intermediate \vred\ of 100--200 \,\kms\, (angle-averaged), as dust preferentially destroys \Lya\ photons that undergo a large number of scatterings. On the other hand, the \CRMHD\ galaxy shows similar \vred\ values because \Lya\ photons propagate preferentially along low $N_\mathrm{HI}$ channels. Orientation effects further increase the scatter (Figure~\ref{fig:BR}-c). Both simulations predict values of $\vsep\sim160\,\kms$ and $\BR\sim0.7$ in the mock spectra, which are broadly consistent with existing observations, although the average $\BR$ values tend to be higher than the observed ones.

\item The simulations post-processed with LyC radiative transfer do not produce galaxies with $\vred\gtrsim350\,\kms$ and $\BR<0.5$. However, when the neutral hydrogen fraction is assumed to be determined by collisional equilibrium and fills the optically thin channels near young stars, the galaxies affected by CRs yield large angle-averaged \vred\ values of up to $400\, \kms$ (Figures~\ref{fig:BR}-a). This simple experiment suggests that galaxies with large $\vred$ may be better explained if more volume-filling neutral hydrogen is present within the inner virial sphere (Figure~\ref{fig:large_vred}).

\item The \CRMHD\ run yields flux ratios between the valley and maximum of the two peaks  $\fvalley=0.02\pm0.02$, consistent with the low-$z$ LSAD sample ($\fvalley=0.04\pm0.07$). The over-cooled galaxy without CR predicts a slightly higher $\fvalley=0.12\pm0.13$.

\item CR feedback produces more extended HI distributions in the CGM (Figure~\ref{fig:img}). The model reproduces the \Lya\ SB of faint LAEs observed with MUSE \citep{wisotzki18} (Figure~\ref{fig:sbp}), while the over-cooled galaxy exhibits slightly fainter profiles. Our results also confirm previous findings that approximately half of the \Lya\ emission in the CGM region likely originates from collisional excitation, while the other half comes from recombinative radiation by young stars.

\item LyC escape fraction (\fLyC) is predicted to be lower with CR feedback, due to the extended \HI\ distribution, although the difference in \fLyC\ between the two simulations is not sufficiently significant to place strong constraints on the feedback models. For non-LyC leakers ($\fLyC\la 0.1$), the peak separation of \Lya\ (\vsep) shows little correlation with \fLyC\ (Figure~\ref{fig:vsep}). 

\end{itemize}

While individual \Lya\ features, such as \BR\ or \vred, may not directly constrain the galaxy formation models, the differences between the simulated and observed mean \BR\ suggest that current simulations still need to improve CR transport \citep[e.g.,][]{hopkins21_CR,nunez-castineyra22}, resolution \citep[e.g.,][]{peeples19,suresh19}, and/or lack crucial baryonic processes, such as bursty star formation \citep[e.g.,][]{kang25}. Comparison of simulated \Lya\ features with statistically significant samples of LAEs remains valuable, offering an independent perspective. Furthermore, numerical modeling of \Lya\ emission at different scales is essential for interpreting statistical LAE data and leveraging these samples to constrain structure formation and cosmology. The sensitive nature of \Lya, rather than being a limitation, provides a unique opportunity for future astrophysical studies.

\section*{acknowledgments}
We thank Lutz Wisotzki for sharing their data with us. TK was supported by the National Research Foundation of Korea (RS-2022-NR070872 and RS-2025-00516961) and also by the Yonsei Fellowship, funded by Lee Youn Jae. HS acknowledges the support of the National Research Foundation of Korea (No. 2022R1A4A3031306), funded by the Korean government (MSIT).  This work used the DiRAC Data Intensive service (DIaL2 / DIaL3) at the University of Leicester, managed by the University of Leicester Research Computing Service on behalf of the STFC DiRAC HPC Facility (www.dirac.ac.uk). The DiRAC service at Leicester was funded by BEIS, UKRI and STFC capital funding and STFC operations grants. DiRAC is part of the UKRI Digital Research Infrastructure. The supercomputing time for numerical experiments was kindly provided by KISTI (KSC-2024-CRE-0200), and large data transfer was supported by KREONET, which is managed and operated by KISTI.

\appendix
\restartappendixnumbering
\renewcommand{\thefigure}{A\arabic{figure}}
\section{Post-processing of simulations}
\label{sec:appendix}
\begin{figure}
	\includegraphics[width=\linewidth]{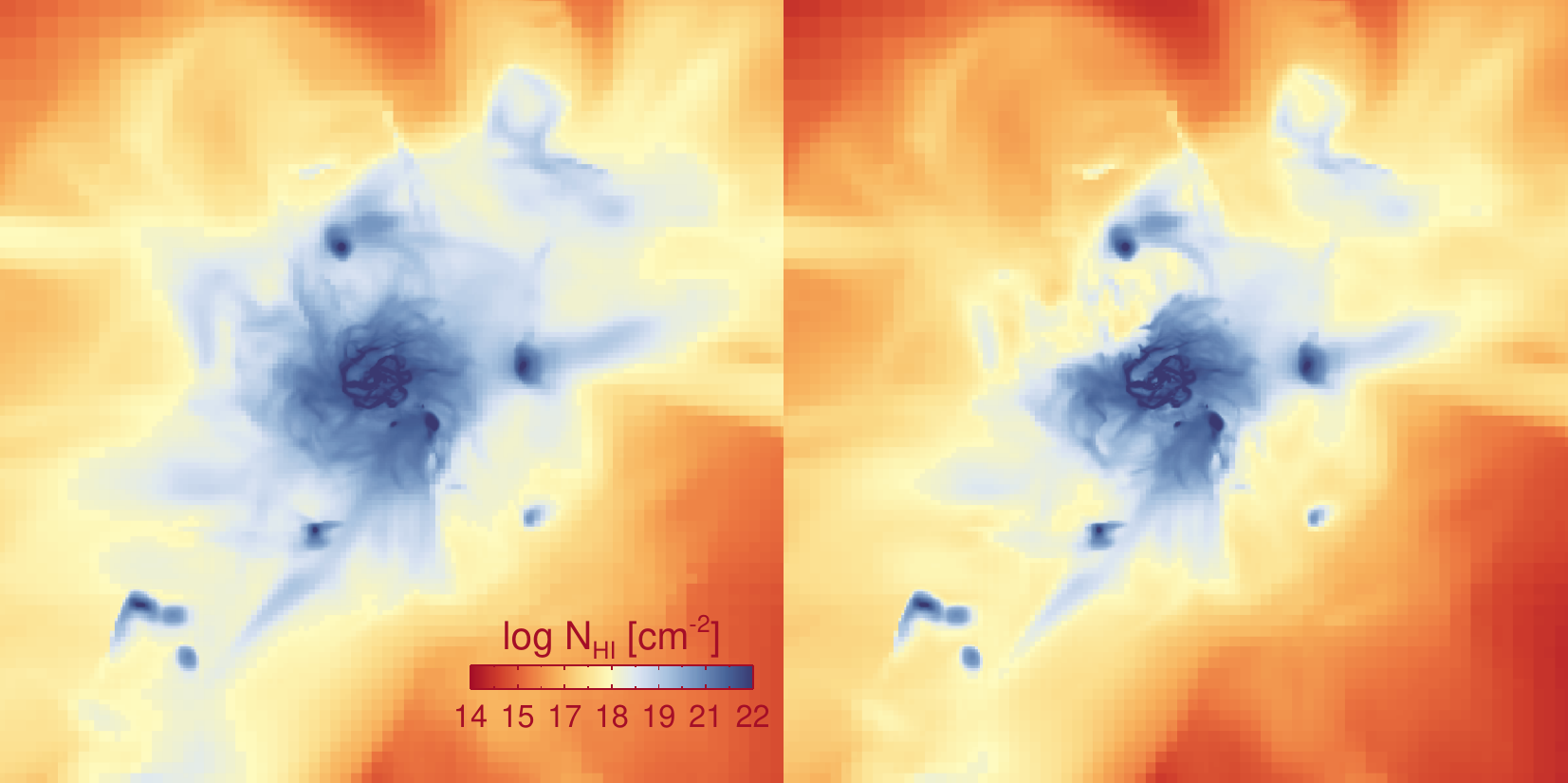}
    \caption{ Example of $N_\mathrm{HI}$ map before (left) and after (right) the LyC post-processing. }
    \label{fig:img_pp}
\end{figure}

We post-process the simulations with ionizing radiative transfer using {\sc ramses-rt} to obtain more accurate estimates of neutral hydrogen. This is accomplished by running the simulations without updating the hydrogen densities but only changing the temperature and ionization fractions. We select three random snapshots and run  {\sc ramses-rt}  for $\sim 50\,\mathrm{Myr}$, determining the timescale for the LyC escape fractions to converge to >98\% (20 Myr). We then post-process all the snapshots for 20 Myr before performing the \Lya\ radiative transfer. Figure~\ref{fig:img_pp} shows an example of the change in ionization structures after the LyC radiative transfer. Most of the neutral hydrogen remains nearly intact, while $N_\mathrm{HI}$ is notably reduced along some channels. On average, $N_\mathrm{HI}$ changes only to a small extent.

\bibliographystyle{aasjournalv7}
\bibliography{main_apj}

\end{document}